\documentclass[twocolumn,twoside,slac]{revtex4}
\usepackage{graphicx}
\usepackage{fancyhdr}
\begin{document}

\title{A New Implementation of the Region-of-Interest Strategy for the
  ATLAS Second Level Trigger}

%

\author{S. Armstrong, V. Boisvert \footnote{Presenter at the conference}, S. Brandt, M. Elsing}
\affiliation{European Laboratory for Particle Physics, CERN, Geneva, Switzerland}
\author{J.T. Baines, W. Li}
\affiliation{Rutherford Appleton Laboratory, Chilton, Didcot, UK}

\author{S. George}
\affiliation{Department of Physics, RHBNC, University of London,
  Egham, UK}

\author{A.G. Mello}
\affiliation{Universidade Federal do Rio de Janeiro, COPPE/EE/IF, Rio
  de Janeiro, Brazil}

\author{\vspace*{0.2cm}
On behalf of the Atlas High Level Trigger Group \cite{authors} }

\begin{abstract}
Among the many challenges presented by the future ATLAS detector at the LHC are the high data taking rate and volume and the derivation of a rapid trigger decision with limited resources. To address this challenge within the ATLAS second level trigger system, a Region-of-Interest mechanism has been adopted which dramatically reduces the relevant fiducial volume necessary to be readout and processed to small regions guided by the hardware-based first level trigger. Software has been developed to allow fast translation between arbitrary geometric regions and identifiers of small collections of the event data. This facilitates on-demand data retrieval and collection building. The system is optimized to minimize the amount of data transferred and unnecessary building of complex objects. Details of the design and implementation are presented along with preliminary performance results. 

\end{abstract}

\maketitle

\thispagestyle{fancy}


\section{INTRODUCTION}

ATLAS (A Toroidal LHC ApparatuS) is one of the four detectors currently
being built around the LHC (Large Hadron Collider) at the European
Organization for Nuclear Research (CERN) in Geneva, Switzerland. The
LHC will collide protons at a centre-of-mass
energy of 14TeV. ATLAS is a multipurpose detector designed to have a
4$\pi$ hermeticity around the interaction region. The physics goals of
ATLAS are numerous: from Higgs searches, to Top physics, from SUSY
searches to QCD physics, not forgetting $B$ physics as well as exotic
searches. In addition, part of the excitement for reaching this energy
frontier is the discovery potential associated with this unchartered
territory. These physics requirements combined with a bunch crossing
rate of 40 MHz in addition to an average of 25 inelastic proton-proton
underlying interactions in each bunch crossing (at a design
luminosity of $10^{34} cm^{-2}s^{-1}$) put very stringent constraints
on the data acquisition and trigger system. 

\begin{figure*}[t]
\centering
\includegraphics[width=135mm]{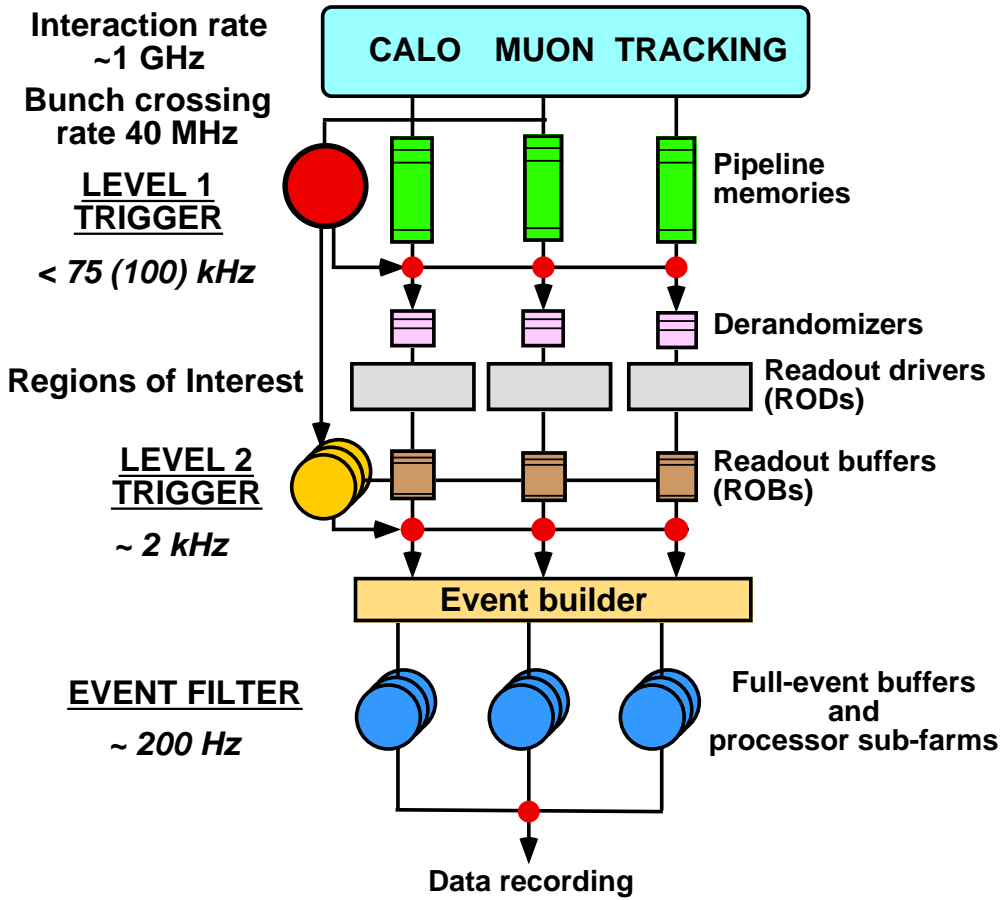}
\caption{Block diagram of the Trigger/DAQ system.} \label{fig:TrigArch}
\end{figure*}

The overall architecture of the three-level ATLAS trigger system is
shown in figure \ref{fig:TrigArch} \cite{TP}. It is designed to reduce the nominal 40 MHz bunch
crossing rate to a rate of about 200 Hz at which events, that will
have a size of about 1.6 MB on average, will be written to mass
storage. The first stage of the trigger, LVL1, is hardware-based and
it reduces the rate to about 75KHz. Using the fast calorimeter and
muon sub-detectors, it has a latency (time taken to form and distribute
the LVL1 trigger decision) of about 2.5 $\mu s$. During that time,
the data from all the sub-detectors (about $10^8$ electronic channels)
are kept in pipeline memories. After the LVL1 decision, selected data
fragments are transferred to the Readout Drivers (RODs) and then to
the Readout Buffers (ROBs). It is foreseen that there will be a total
of about 1700 ROBs. The second stage of the trigger system, LVL2, is
software-based and it reduces the rate to about 2KHz. Making use of
the so called Region-of-Interest mechanism the average latency is
about 10ms. In order to achieve this goal, the main characteristic of
this stage is a fast rejection achieved by optimized trigger
algorithms. The last stage of the trigger system, the Event Filter,
occurs after the event building process. At this stage, the average
latency is about 1s. The goal of the Event Filter is both to reduce
the rate to about 200Hz, necessary for mass storage, but also to
classify the events. Hence full calibration and alignment information
is available at this stage. The trigger algorithms used for this stage
have much in common with the offline algorithms and massive reuse of
those is foreseen. The LVL2 and the Event Filter stages are commonly
referred to as the High Level Trigger (HLT).

\subsection{The Region-of-Interest Mechanism}

\begin{figure}[t]
\centering
\includegraphics[width=65mm]{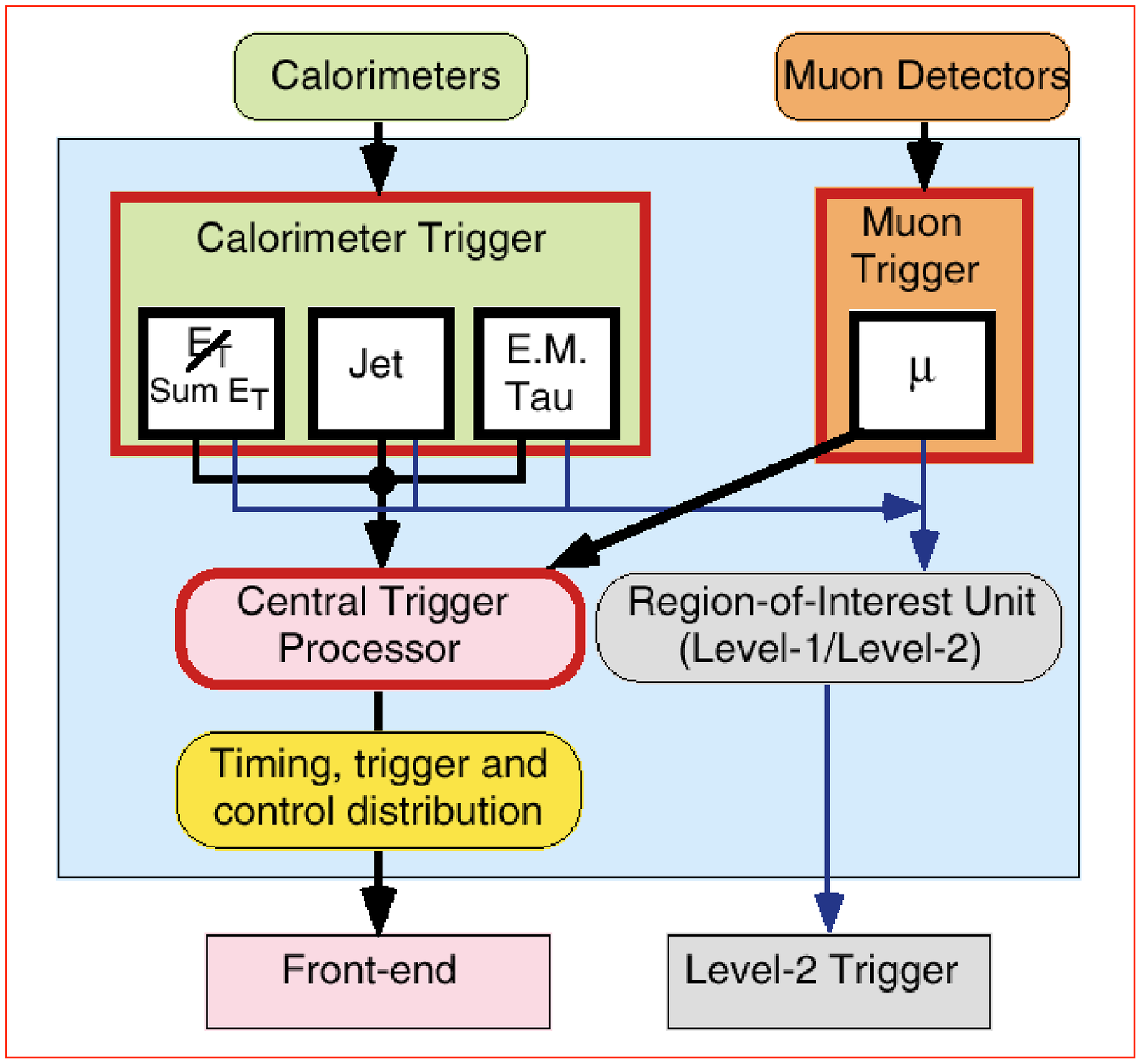}
\caption{Block diagram of the Region-of-Interest mechanism.} \label{fig:Roi}
\end{figure}

An important piece of the strategy of the ATLAS trigger relies on the
Region-of-Interest (RoI) mechanism for which the LVL2 trigger makes
use of information provided by the LVL1 trigger in localized region of
the calorimeter and muon sub-detectors. This process is shown
schematically in figure \ref{fig:Roi}. The information contained in the RoI
typically include the position ($\eta$ and $\phi$) and the $p_T$ of
the candidate objects as well as energy sums. Candidate objects
selected by the LVL1 can be high-$p_T$ muons, electrons or photons,
hadrons or taus, and jets. The energy sums include the missing-$E_T$
vector and the scalar $E_T$ value. For all selected LVL1 events, the
RoI information is sent to the LVL2 using a dedicated data
path. Making use of this RoI information, the LVL2 algorithms only
transfer the necessary ROBs in order to arrive quickly at a LVL2 decision. It is
important to note that all the data from all the sub-detectors with
full granularity is available for the LVL2 algorithms if
necessary. However, typically only a small fraction of the detector,
centred around the RoI information selected by the LVL1, is needed by
the LVL2 algorithms. On average there are a few RoI per event
and as a consequence to this mechanism only a few percent of the total
event data is required at the LVL2 stage. On the other hand, even
though the bandwidth requirements on the Data Acquisition system is
much reduced, it makes for a more complex system.

\subsection{The Data Access}
\label{sec:dataacc}

\begin{figure*}[t]
\centering
\includegraphics[width=135mm]{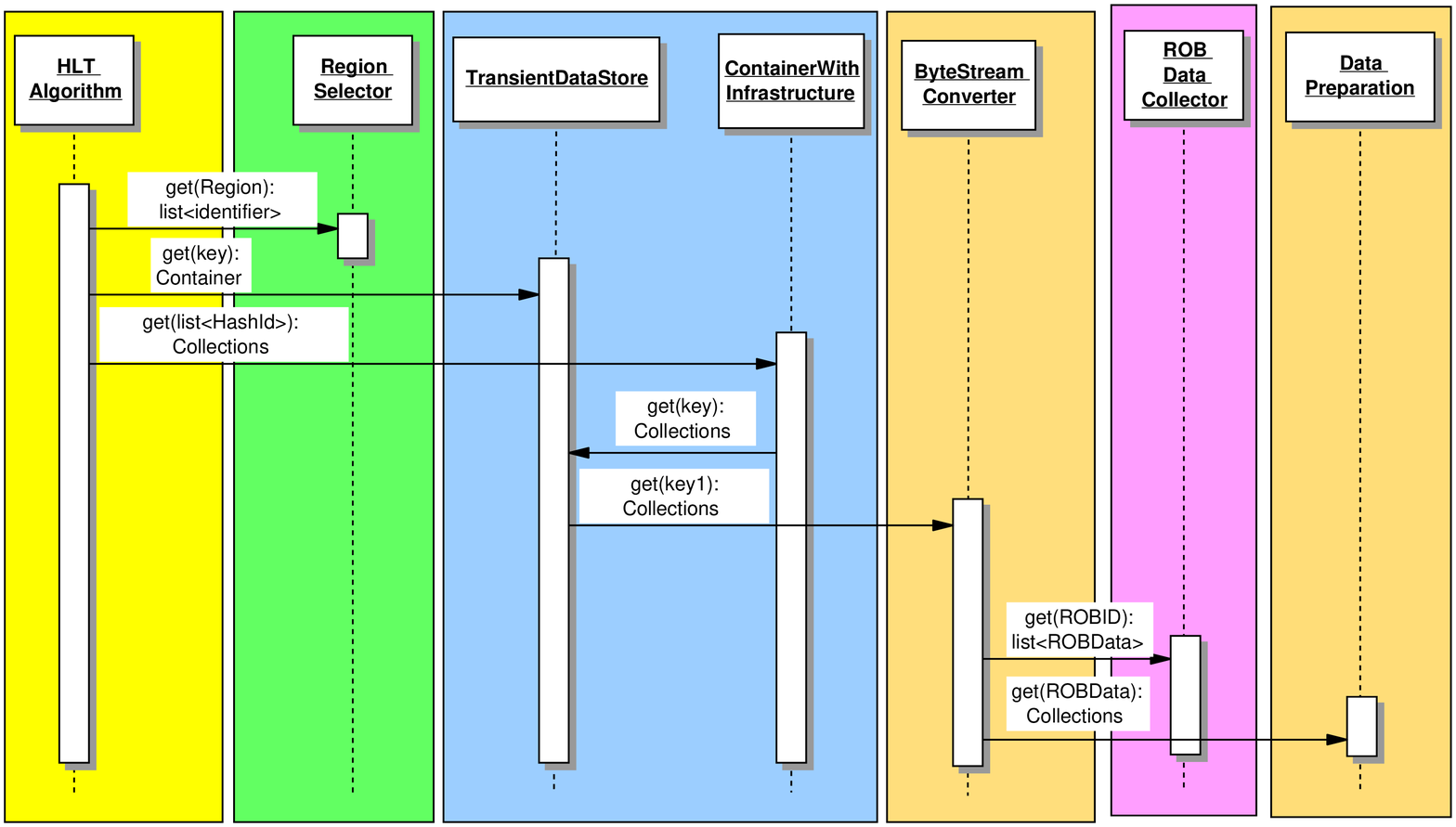}
\caption{Schematic sequence diagram of the data access.} \label{fig:london}
\end{figure*}

As discussed above, the Region-of-Interest mechanism allow for an
optimized retrieve of the data necessary to perform the LVL2
decision. Figure \ref{fig:london} shows in more details the sequence
associated with the data access. The data is stored in collections
within the Transient Data Store (TDS) \cite{paolo}. These collections are organized
inside a ContainerWithInfrastructure for easy retrieval and
sorting. The collections are identified uniquely via an offline
identifier \cite{offlineid}. We
can see from the diagram that the HLT algorithm first ask the
RegionSelector tool for the list of collection identifiers associated
with a particular region, that could correspond for example to an
$\eta \phi$ region that would come from LVL1. With the list of
collection identifiers in hand, the HLT algorithm request the
associated data to the TDS. If the data is already cached within the
TDS, the requested collections are returned. If the data is not
cached, the TDS launches the ByteStreamConverter which goal is to fill
the collections in the TDS using the data in ByteStream format. To get
a hold of this data, the ByteStreamConverter must request specific
ROBs to the ROBDataCollector. Finally, at LVL2 there is the
possibility to do some data preparation from within the
ByteStreamConverter, which leads to a faster execution of the LVL2
trigger algorithms. One consequence from this sequence is the fact
that the RegionSelector tool plays a central role within the trigger
chain, since every Trigger Algorithm that needs access to the data in
a certain region will have access to this tool. Another remark that
stems from this data access sequence is the need for an optimization
of the collection granularity. There needs to be a trade off between a
useful navigation for the trigger algorithms and a minimization of the
data requests. Finally, we mentioned that the collection request was
made using the offline identifiers, rather than the online
identifiers. This choice comes primarily from two issues, one is the
fact that the current design for the trigger architecture calls for
the use of offline code in the online environment. That is to say,
both the trigger algorithms and the architecture in which they run can
be developed in an offline environment and be directly ported to the
online environment. The benefits are numerous: this will facilitate
the development of algorithms; this will allow the study of the
boundary between LVL2 and Event Filter and it will lead to easy
performance studies for physics analysis. The second issue related
with the use of offline identifiers for the collection and the
RegionSelector comes from the fact that a possible region for which
a trigger algorithm could require data is the InnerDetector
sub-detectors, for which there are no LVL1 online identifiers. 

\section{THE REGION SELECTOR TOOL}
Having described the usefulness and environment surrounding the
RegionSelector tool we now turn to its requirements and implementation.
\subsection{Requirements}

As we have just seen, the RegionSelector tool is central to any LVL2
trigger algorithm, hence its foremost requirement is that it should be
fast and use up only a fraction of the available latency at
LVL2. Another requirement imposed on the RegionSelector is the fact
that it should translate an arbitrary geometrical region into a list
of collection identifiers. Such a region can be a simple cone that
span the various sub-detectors. It can also be a more complex cone
which accounts for the uncertainty in the $z$ position of the primary
vertex, coming from the beam spread. In that case, $\Delta \eta$ has a
radial dependence. Finally, another geometrical region of interest is
that of a helical road, which could correspond for example to a
reconstructed track in need of confirmation by a more refined Event
Filter algorithm which has access to calibration and alignment
constants. In the current implementation of the RegionSelector tool
the innermost sub-detectors take into account the $z$ direction spread
while the outermost sub-detectors follow a cone.

\subsection{Implementation}
\begin{figure*}[t]
\centering
\includegraphics[width=135mm]{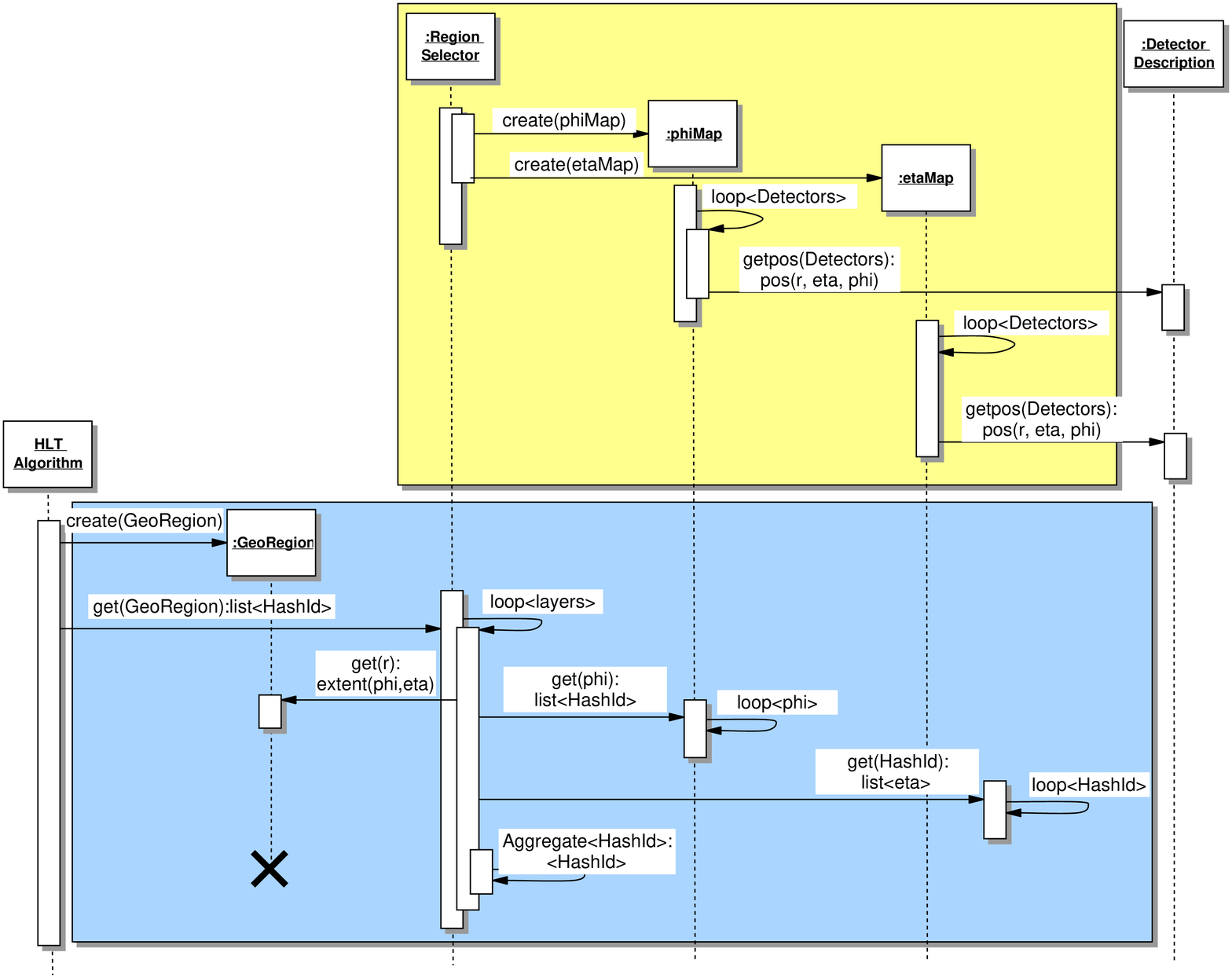}
\caption{Sequence diagram of the Region Selector.} \label{fig:regsel}
\end{figure*}

\begin{figure}[t]
\centering
\includegraphics[width=65mm]{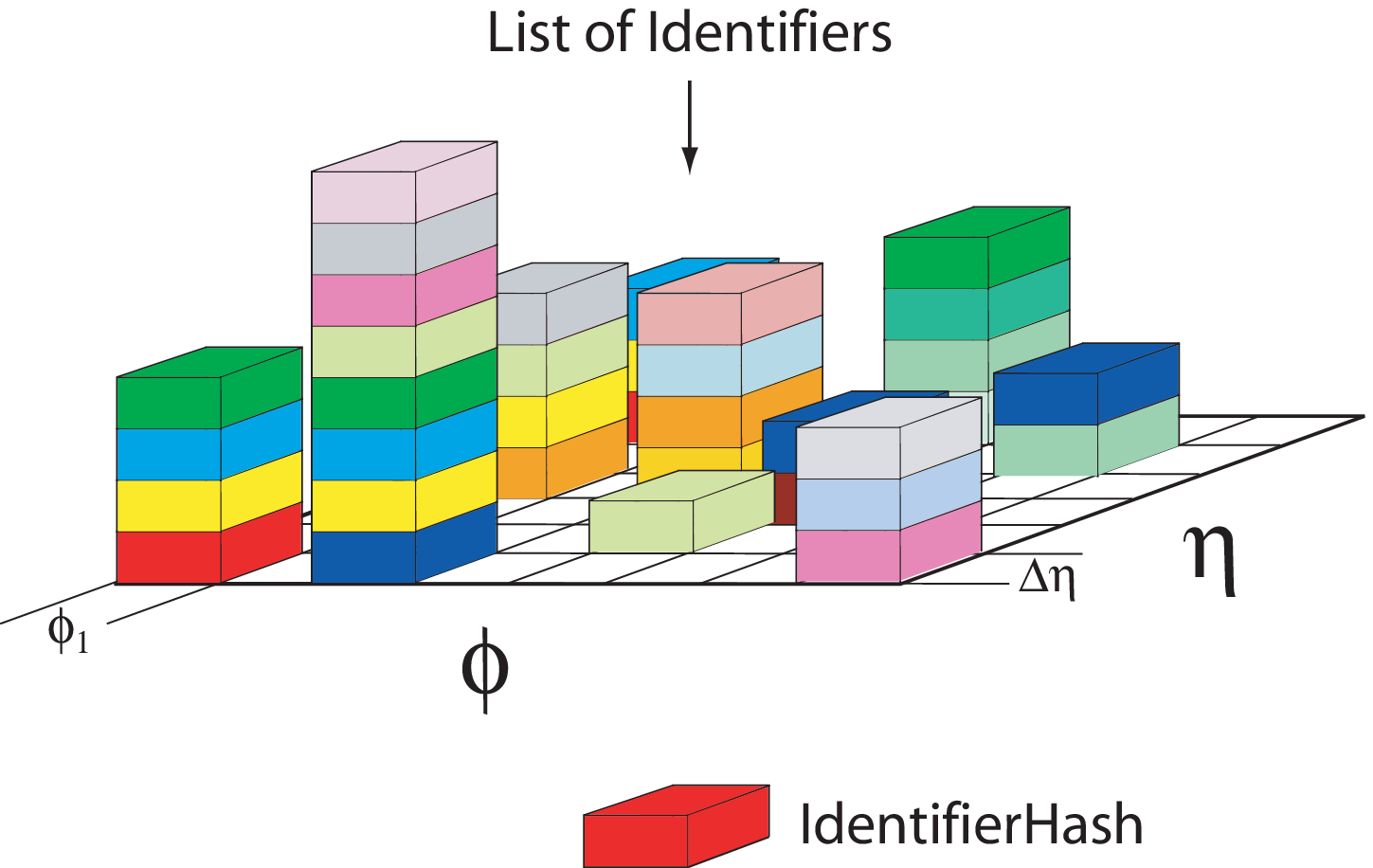}
\caption{Diagram showing schematically the $\eta$ $\phi$ map.} \label{fig:etaphimap}
\end{figure}

In figure \ref{fig:regsel} we show the sequence diagram associated
with the RegionSelector tool. We can distinguish two main parts, the
top one represents the initialization phase, and the lower part shows
the execution phase. At initialization two maps are filled for each
layer of each sub-detector: a map between the $\phi$ index and a set of identifiers, and a
second map between an identifier and a vector of a range in
$\eta$. Figure \ref{fig:etaphimap} shows schematically the concept of
a set of identifiers corresponding to a value in $\phi$ and a range
in $\eta$. During the execution phase, the algorithm asks for the list of
identifiers corresponding to a range in $\phi$ and a range in
$\eta$, for a specified sub-detector. There is an outer loop over the
layers, then there is a loop over the $\phi$ range to get the
associated sets of identifiers. Following this step there is also a
loop over the identifiers to check if the $\eta$ value is within the
required range. The built identifier list is then returned to the
algorithm. This procedure ensures that there are no duplicate
identifier within the list. Care must be taken regarding the $\phi$
compact boundary.  

A key ingredient to the RegionSelector tool are mappings made by the
sub-detector communities that tie together a range in $\eta$ and
$\phi$ and a collection identifier. Those mappings are then used to
make the internal maps during the initialization period of the
RegionSelector. As mentioned in section \ref{sec:dataacc}, there needs
to be an optimization performed for each sub-detector concerning the
granularity of the collections. In the current implementation table \ref{tab:coll}
shows the current collection granularity for each sub-detector, as
well as the total number of ROBs for this sub-detector. For the Pixel
sub-detector a collection corresponds to a module, which is a single
silicon wafer. For the silicon strip detector (SCT) the collection
granularity corresponds to a side of a module which is a bonded pair
of wafers whose strips are oriented in the same direction, either
axial or stereo. For the Transition Radiation Tracker (TRT)
sub-detector, the collection granularity corresponds to a radial straw
layer in the barrel and to 1/32 in $r\phi$ in the endcap wheels. For
the Liquid Argon calorimeter the collection granularity is that of a
Trigger Tower while for the barrel hadronic Tile calorimeter the collection
granularity is a phi wedge. Finally for the muon spectrometer there are four technologies
used. Two types of chambers are used for trigger purposes: the
Resistive Plate Chambers (RPC) and the Thin Gap Chambers (TGC); and
two types of chambers are used for precision measurements: the
Monitored Drift Tube chambers (MDT) and the Cathode Strip Chambers
(CSC). For all of the muon spectrometer the collection granularity
corresponds to a single chamber. For more information on each
sub-detector see for example \cite{atlas}.

\begin{table}[t]
\begin{center}
\caption{Current collection granularity and number of ROBs for each sub-detector.}
\begin{tabular}{|l|c|c|c|}\hline 
 & \textbf{Collection} & \textbf{Number} & \textbf{Num. ROBs} \\
  \hline 
Pixel & module & 1744 & 81 \\
SCT & side of module & 8176 & 256 \\
TRT & straw layer & 19008 & 256 \\
LAr & Trigger Tower & 7168& 768 \\
Tile & module & 256 & 32 \\
muon MDT & chamber & 1168 & 192 \\
muon CSC & chamber & 32 & 32 \\
muon RPC & chamber & 574 & 32 \\
muon TGC & chamber & 1584 & 32 \\ \hline
\end{tabular}
\label{tab:coll}
\end{center}
\end{table}

\section{TIMING MEASUREMENTS}

As mentioned earlier, the main requirement for the RegionSelector tool
is that it should only use up a small fraction of the available latency
at LVL2. Preliminary timing measurements were performed on a 1GHz
Pentium III machine using the TAU (Tuning and Analysis Utilities) \cite{TAU}
timing tool. Table \ref{tab:regsel} shows the timing measurements for various
sub-detectors for different ranges in $\eta$ and $\phi$. 

\begin{table}[t]
\begin{center}
\caption{Timing measurements of RegionSelector in ms.}
\begin{tabular}{|c|l|c|c|c|}\hline 
 && \multicolumn{3}{|c|}{$\Delta \eta$} \\ \hline
 $\Delta \phi$ && \textbf{0.1} & \textbf{0.2} & \textbf{0.5} \\
  \hline 
&Pixel ($\sigma$ = 0.06) & 0.20 & 0.22 & 0.23 \\
&SCT ($\sigma$ = 0.11) & 0.56 & 0.59 & 0.62 \\
&TRT ($\sigma$ = 0.23) & 1.05 & 1.12 & 1.21 \\
0.1 &LAr ($\sigma$ = 0.06) & 0.33 & 0.33 & 0.35 \\
&Tile ($\sigma$ = 0.008) & 0.03 & 0.03 & 0.03 \\
&MDT ($\sigma$ = 0.038) & 0.06 & 0.06 & 0.06 \\
&RPC ($\sigma$ = 0.009) & 0.05 & 0.05 & 0.06 \\ \hline
&Pixel & 0.22 & 0.22 & 0.23 \\
&SCT & 0.60 & 0.61 & 0.63 \\
&TRT  & 1.13 & 1.15 & 1.23 \\
0.5&LAr  & 0.33 & 0.34 & 0.35 \\
&Tile  & 0.03 & 0.03 & 0.03 \\
&MDT  & 0.06 & 0.06 & 0.07 \\
&RPC  & 0.05 & 0.05 & 0.06 \\ \hline
\end{tabular}
\label{tab:regsel}
\end{center}
\end{table}

We can see from the results that the timing is mainly independent on
the extent of the $\eta$ and $\phi$ range. If one looks at the LAr
calorimer timing and arbitrarily divide by 3 to account for a 4GHz
machine, one gets a timing of the Region Selector of about
0.11ms. Since preliminary timing measurements of LVL2 calorimeter
algorithms are of the order of 1ms on a 1GHz machine, we get a
combined RegionSelector and algorithm timing of less than 0.5ms,
extrapolated to a 4GHz machine. This number is well below the expected
10ms average latency allowed at LVL2. Note that the timing for the
data access is not included in this number.

\section{CONCLUSION}

We have shown an implementation of a tool used to translate a region
into a list of collection identifiers. This tool is used by all
algorithms that request data access in a given region of the ATLAS
detector. Although this tool is of crucial importance at LVL2,
specially when combined with the Region-of-Interest mechanism, it can
also be of used in the Event Filter and in offline reconstruction
whenever an algorithm is interested in a particular seed, be it a
simple sub-detector region or an object spanning a sub-detector
region. Already the timing budget of this tool is within the accepted
latency, scaled to a representative machine speed for 2007. It is
foreseen that continuous improvements will be achieved in the
implementation and that more complex geometrical regions will be supported.

\begin{acknowledgments}
The authors wish to thank the ATLAS PESA core software group. The
authors also wish to thank the various sub-detector communities for
their assistance in providing the necessary mappings, central to the
implementation of the RegionSelector tool: K. Assamagan, S. Goldfarb,
G. Gorfine, F. Luehring, H. Ma, S. Sivoklokov, and S. Solodkov. 
\end{acknowledgments}


\end{document}